\documentclass[aps,prd,twocolumn,nofootinbib,showpacs]{revtex4}
\pdfoutput=1

\usepackage{epsf,epsfig,amsmath,amssymb,dsfont}
\usepackage{amsthm,mathrsfs} 
\usepackage{graphicx} 
\usepackage{multirow}
\usepackage{float}
\usepackage{colortbl}
\usepackage[table]{xcolor}
\usepackage{varioref}
\usepackage{slashed}
\usepackage{float}
\usepackage{hyperref}

\newcommand{\be}{\begin{equation}}
\newcommand{\ee}{\end{equation}}
\newcommand{\baln}{\begin{align}}
\newcommand{\ealn}{\end{align}}
\newcommand{\ben}{\begin{equation*}}
\newcommand{\een}{\end{equation*}}

\long\def\symbolfootnote[#1]#2{\begingroup%
\def\thefootnote{\fnsymbol{footnote}}\footnote[#1]{#2}\endgroup}

\newcommand{\nn}{\nonumber \\}

\newcommand{\fr}{\frac}
\newcommand{\del}{\partial}
\newcommand\CS{\mathcal{C}}

\newcommand{\kvec}{\bold{k}}

\begin{document}

\title{Spectral Dimension from Nonlocal Dynamics on Causal Sets}

\author{Alessio Belenchia, Dionigi M. T. Benincasa}
\affiliation{SISSA - International School for Advanced Studies, Via Bonomea 265, 34136 Trieste, Italy.}
\affiliation{INFN, Sezione di Trieste, Trieste, Italy.}
\author{Antonino Marcian\`o and Leonardo Modesto}
\affiliation{Department of Physics \& Center for Field Theory and Particle Physics, Fudan University, 200433 Shanghai, China.}


\begin{abstract}

We investigate the spectral dimension obtained from non-local continuum d'Alembertians derived from causal sets. 
We find a universal dimensional reduction to 2 dimensions, in all dimensions. We conclude by discussing the validity and
relevance of our results within the broader context of quantum field theories based on these nonlocal dynamics.

\end{abstract}

\maketitle
\flushbottom

\section{Introduction}\label{introduction}

In recent years, the study of the spectral dimension as a tool to investigate the small scale structure of spacetime has attracted much attention. Indeed, analyses performed in causal dynamical triangulations \cite{Ambjorn:2005db}, asymptotically safe quantum gravity \cite{Lauscher:2005qz}, loop quantum gravity \cite{Modesto:2008jz}, super-renormalizable gravity \cite{Modesto:2011kw,Modesto:2012ys,Modesto:2014lga}, noncommutative geometry \cite{Benedetti:2008gu,Arzano:2014jfa} and in cosmology \cite{Calcagni:2009ar}, all point towards a running of the spectral dimension at short scales \cite{Calcagni:2014wba}. 
In particular, dimensional reduction appears to be ubiquitous in quantum gravity (see \cite{Carlip:2009km}). 

Although several approaches to quantum gravity share the property of dimensional reduction on small scales \cite{Carlip:2009km}, the causal set approach is an outlier. Indeed, results by Eichhorn and Mizera \cite{Eichhorn:2013ova}  (EM from here on) show that the spectral dimension of a causal set increases at small scales. Their analysis relies on computing the spectral dimension from the return probability, $P$, of a random walker on a causal set with the spectral dimension\footnote{They also defined a ``causal" spectral dimension using the meeting probability of two random walkers, $P_M$, both moving forward in time, which give similar results.} defined as $d_{EM}=-2\del\ln P(s)/\del\ln s$. They then argue that the increasing behaviour of $d_{EM}$ at short scales is due to the radical nonlocality present in causal sets \cite{Benincasa:2010ac}. This exact same non-locality which, as we will see shortly, arises because of the interplay between Lorentz invariance and discreteness, also appears in the construction of d'Alembertians governing the dynamics of scalar fields on causal sets \cite{Sorkin:2007qi}. An obvious question then is whether the spectral dimension computed from such d'Alembertians using the usual heat kernel techniques, corroborates the EM result.

The rest of the paper is organised as follows. In Section 2 we briefly introduce causal sets, the notion of sprinklings and how the non-locality referred to above arises within this context. We then introduce the family of d'Alembertians defined in \cite{Aslanbeigi:2014tg} 
in both position and Fourier space, and discuss some of their properties. Section 3 is devoted to computing the spectral dimension of  Minkowski spacetime. In particular, we determine both the short and large scale behaviour of the spectral dimension in all dimensions analytically, and for particular choices of nonlocal d'Alembertians in 2, 3 and 4 dimensions we numerically compute the full dependence of the spectral dimension on the diffusion parameter. In Section 4 we discuss our results, and in Section 5 we spell out some conclusions.

\section{Causal sets, Non-locality and d'Alembertians}\label{introduction}

Before we define the d'Alembertians that will be used in calculating the spectral dimension we briefly review how the non-locality referred to in the introduction arises. 

\subsection{Causal Set Nonlocality} 

Recall first that a causal set (causet) is a locally finite partially ordered set, where the partial order is taken to underly the macroscopic causal order of spacetime events~\cite{Bombelli:1987aa}. In order to unite fundamental discreteness with Lorentz invariance a kinematic randomness must be introduced. This is done via the notion of a {\it sprinkling}, i.e. a Poisson process of selecting points in a spacetime $\mathcal{M}$ uniformly at random at some density $\rho=1/l^d$, such that the expected number of points in a spacetime region of volume $V$ is $\rho V$. This process generates a causet, $\CS$, the elements of which are the sprinkled points and the order relation of which is the spacetime's causal relations restricted to the sprinkled points. The randomness of the sprinkling is key in ensuring that the causet thus generated is Lorentz invariant \cite{Bombelli:2006nm}, and it leads to the following radical form of non-locality. Consider a causet $\CS$ that is a sprinkling of $d$ dimensional Minkowski spacetime, and let $p\in\CS$. Then the nearest neighbours to, say the past of, $p$ will be the set of points $q$ past-linked to $p$, i.e. $\{ q\prec p\; |\; \nexists \,r \;\text{s.t.} \; q\prec r\prec p\}$. The locus of such points will roughly comprise the space-like hyperboloid, $\Sigma$, to the past of $p$ defined by $\Sigma=\{q\in J^-(p)\;|\; \tau(p,q) = l\}$. Hence there will be an infinite number of such points. Precisely this form of non-locality is evident in the definition of d'Alembertian operator on a causal set, defined by constructing a finite difference equation in which linear combinations of the value of the field at neighbouring points are taken. Since the number of nearest neighbours, next nearest neighbours, etc., is infinite, the corresponding expression looks highly non-local (see \cite{Benincasa:2010ac} and \cite{Sorkin:2007qi}).

\subsection{Nonlocal d'Alembertians}

Recently, Aslanbeigi {\it et.~al} (\cite{Aslanbeigi:2014tg}) generalised the original constructions of nonlocal d'Alembertians in~\cite{Sorkin:2007qi,Benincasa:2010ac,Dowker:2013vl}, to include an infinite family of nonlocal d'Alembertians for any dimension $d$. The nonlocal d'Alembertians $\Box^{(d)}_\rho$, labelled by a non-locality scale $\rho$ --- originally taken to coincide with the fundamental discreteness scale of causal sets, but need not be --- which we take to be a free parameter of the theory, are given by
\begin{widetext}
\be
\Box_{\rho}^{(d)}\phi(x) = \rho^{\fr{2}{d}}\left(\alpha^{(d)}\phi(x) +\rho\, \beta^{(d)}\sum_{n=0}^{N_d}C_n^{(d)}\int_{J^-(x)}d^dy \,\fr{(\rho V(x,y))^n}{n!}e^{-\rho V(x,y)}\phi(y)\right),
\label{boxd}
\ee
\end{widetext}
where $N_{d}$ is a dimension dependent positive integer, $\rho=1/l_k^d$ 
and the coefficients $\alpha^{(d)}$, $\beta^{(d)}$ and $C_n^{(d)}$ can be 
found in Equations (12)-(15) of \cite{Glaser:2013sf}. 
It is straightforward to recover the original results of \cite{Sorkin:2007qi} in $d=2,4$ as 
the simplest cases of the analysis reported in \cite{Aslanbeigi:2014tg}. 

The momentum space representation of these expressions, $g^{(d)}_\rho(k^2)$, 
were also computed in \cite{Aslanbeigi:2014tg}:
\begin{widetext}
\be
g^{(d)}(k^2)=\rho^{2/d}\left(\alpha^{(d)}+\beta^{(d)}2(2\pi)^{d/2-1}Z^{\fr{2-d}{4}}\sum_{n=0}^{N_d}\fr{C_n^{(d)}}{n!}\gamma_d^n
\int_0^{\infty}ds\,s^{d(n+1/2)}e^{-\gamma_d s^d}K_{\fr{d}{2}-1}(Z^{1/2}s)\right),
\label{dalembFT}
\ee
\end{widetext}
where 
\be
Z=\fr{k\cdot k}{\rho^{2/d}},\quad\qquad \gamma_d=\fr{(\fr{\pi}{4})^{\fr{d-1}{2}}}{d\,\Gamma\left(\fr{d+1}{2}\right)},
\ee
and $K_{\nu}$ is the modified Bessel function of the second kind, possessing a cut along the negative real axis and 
for which we assume the principal value. 

The functions \eqref{dalembFT} simplify considerably in the limits
where $k^2\ll \rho^{2/d}$ and $k^2\gg \rho^{2/d}$, i.e. the infrared (IR) and ultraviolet (UV) limit respectively.
By construction the IR behaviour is the same for all operators, i.e. $g^{(d)}_\rho(k^2)\rightarrow -k^2$ as 
$k^2\rightarrow 0$. The UV limit on the other hand is dimension dependent, and can be shown to be
\be
g_\rho^{(d)}(k^2) \rightarrow a\rho^{2/d} + b\rho^{\fr{2}{d} +1}(k^2)^{-d/2} +\dots,
\ee
as $k\rightarrow\infty$,
where $a$ and $b$ are dimension dependent constants. 
Note that since $g^{(d)}_\rho$ goes to a constant in the UV, the Green 
function \eqref{green} possesses a delta function divergence in the coincidence limit. 
This divergence can be regularised by subtracting the constant $(a\rho^{2/d})^{-1}$
from the momentum space Green function \cite{Aslanbeigi:2014tg}. Inverting back gives
a {\em regularised} momentum space d'Alembertian
%
\be
g^{(d)}_{reg}:=\frac{a \rho ^{2/d} g_{\rho }^d}{a \rho ^{2/d}-g_{\rho }^d},
\label{greg}
\ee
which now has the following UV behaviour  
\be
g^{(d)}_{reg}\rightarrow -\fr{a^2}{b}\rho^{\fr{2}{d}-1}(k^2)^{\fr{d}{2}}+\dots.
\ee
Note that this regularisation procedure is manifestly Lorentzian and is physically
motivated by the underlying theory being a theory on the {\em discrete} causal set, where
a coincidence limit does not exist.

\begin{figure}[H]
\centering
\includegraphics[scale=0.3]{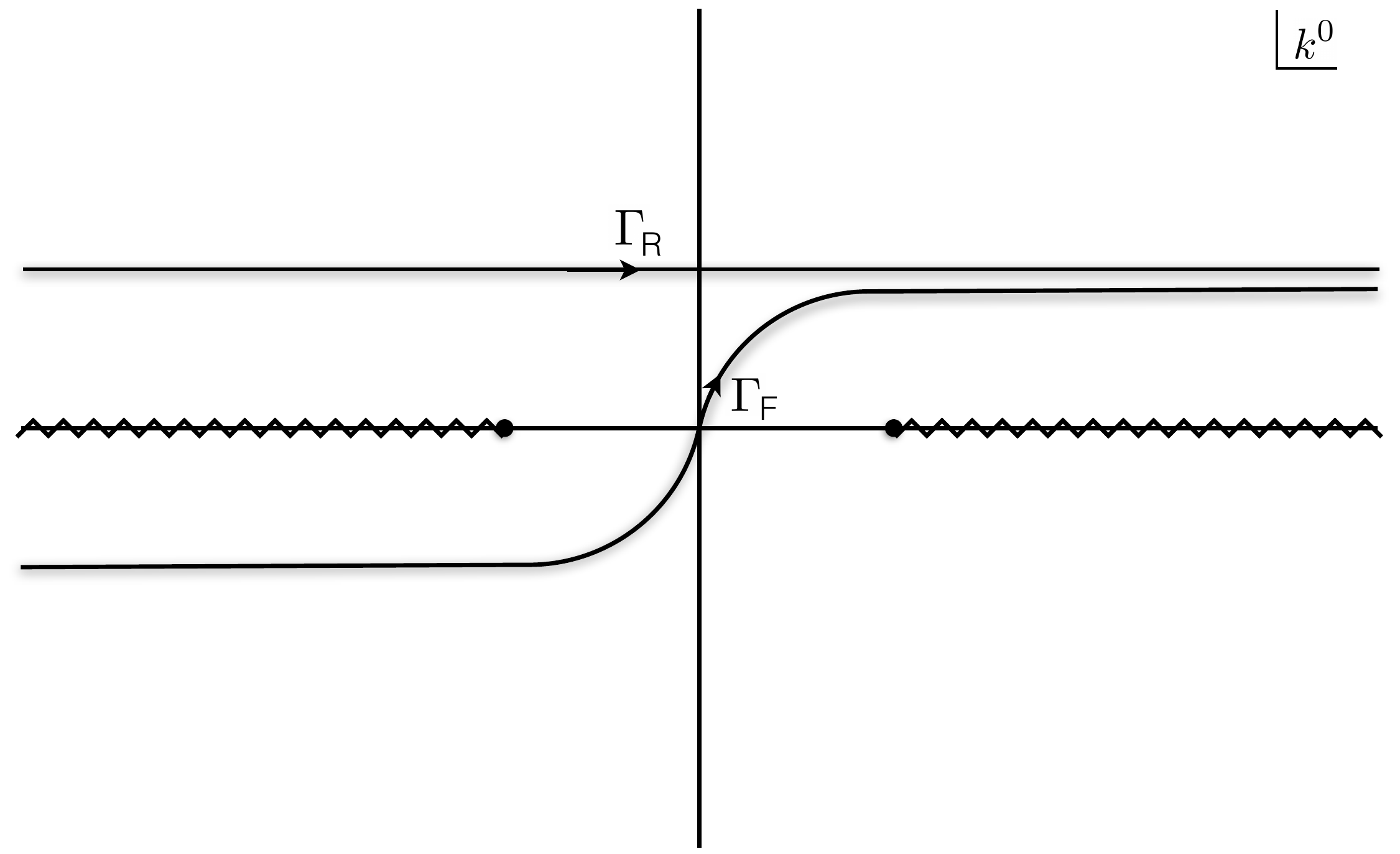}
\caption{A schematic diagram of the singular structure of $1/g(k^2)$ in the complex $k^0$ plane, including
two contours, $\Gamma_R$ and $\Gamma_F$, which define the retarded and Feynman propagators
respectively. Note that before we are able to define the latter we must first analytically continue $g(k^2)$
to the lower half the plane. The contour $\Gamma_R$ will clearly encounter an obstruction under 
a Wick rotation, $k^0\rightarrow -ik^0$, while $\Gamma_F$ won't.}\label{contour}
\end{figure}

The operators we started with \eqref{boxd} are {\em retarded} Lorentz invariant operators, and their
Laplace transforms are therefore defined in the limit $\Im(k^0)\rightarrow 0^+$~\cite{dominguez1979laplace,Aslanbeigi:2014tg,Saravani:2015aa},
i.e. in the upper half complex $k^0$-plane.
Their inverse Fourier transforms therefore yield the (unique) retarded Green functions 
to \eqref{boxd} and are given by
the integral
\be
G^{(d)}(x,y) = \int_{\Gamma_R}\fr{dk^0}{2\pi}\int \fr{d^{d-1}\kvec}{(2\pi)^{d-1}}\fr{e^{ik\cdot(x-y)}}{g^{(d)}(k^2)},
\label{green}
\ee
where $\Gamma_R$ is a contour running from $-\infty$ to $\infty$ in the upper half complex $k^0$ 
plane such that all singularities of $g^{-1}$ lie below the contour, see Fig. \ref{contour}.

In the next section, in order to compute the spectral dimension using heat-kernel
techniques, we will need to to Wick rotate the d'Alembertian, $g_{reg}$, or equivalently
its retarded propagator. 
However such a Wick rotation cannot  be performed on 
the retarded propagator because the contour, $\Gamma _R$, would cross singularities. 
To avoid this issue one must use the 
Feynman propagator whose contour
can be Wick rotated without crossing any singularities (see Fig. \ref{contour}). To define
this propagator we first analytically continue $g$ to the whole complex plane and then define
the Feynman propagator to be \eqref{green} with $g_{reg}$ replaced by its analytically
continued version and $\Gamma_R$ replaced by $\Gamma_F$. Although this propagator is 
not a Green function of the original retarded d'Alembertians, 
which indeed only admit unique retarded propagators, two independent 
studies of free scalar QFTs based on such nonlocal dynamics suggest that it is
the correct Feynman propagator for such theories \cite{Belenchia:2014fda,Saravani:2015aa}.
%
%
%
%
From here on we drop the explicit $\rho$ dependence.

\section{Spectral dimension}\label{ds}

In this section we provide both numerical and analytical results for the spectral dimension computed from the regularised Laplace transform of the non-local d'Alembertian $g^{(d)}_{reg}$. 
The numerical analysis will be restricted to the minimal cases in 2, 3 and 4 spacetime dimensions $d$. 

\subsection{Analytical results}

We first recall that the spectral dimension is defined as 
\be 
d_{s}=-2\frac{\partial \ln(P(s))}{\partial \ln(s)},
\label{spectral}
\ee    
where $P(s):=\text{Tr}(P(s;x,y))$ 
is the return probability for a particle, the diffusion process of which is dictated by the Laplacian of interest via the
equation
\be
\fr{\del}{\del s}P(s;x,y) + \Delta P(s;x,y) =0.
\ee 
Therefore, given a modified d'Alembertian, one has to first find the corresponding Laplacian in Euclidean signature, and then consider the heat equation, on the manifold of interest, determined by the Laplacian at hand \cite{Benedetti:2009ge}.  Equation (\ref{spectral}) can be interpreted as describing diffusion of a fictitious particle on the euclideanised spacetime, the solutions of which are given by the heat kernel. The latter, denoted as $P(s;x,y)$, physically represents the probability density of diffusion from $x$ to $y$ in 
time $s$. The trace of $P$ then gives the return probability for the diffusing particle. 
In flat $d$-dimensional Minkowski spacetime with standard local d'Alembertian for example, the spectral dimension coincides with the 
Hausdorff dimension $d$ of the spacetime, for all diffusion times. 

Consider now a free massless scalar field in $d$-dimensional Minkowski spacetime, with dynamics defined by a member of the regularised nonlocal d'Alembertians, as introduced in Section \ref{introduction}, namely $g^{(d)}_{reg}$. The trace of the heat kernel, $P(s)$, is given by 
\be 
P(s)=\int \frac{d^d k}{(2\pi)^{d}} e^{s \,g_{reg}^{(d)}}= C_d\int _{0}^{\infty}dk \,k^{d-1}\, e^{s\, g_{reg}^{(d)}}\,,
\label{prob}
\ee
where $C_d$ is a dimension dependent constant, and we have analytically continued to Euclidean signature.

The IR behaviour of $g_{\rm reg}^{(d)}\rightarrow -k^2$, as $k^2\rightarrow 0$, ensures that for large diffusion times the spectral dimension will flow to 
the value of the Hausdorff dimension $d$. In the UV instead we have that $g_{reg}^{(d)}\approx k^{d}$, so that the trace of the heat kernel can be written as 
\be \label{ids}
\int_{\lambda}^{\infty}dk\, k^{d-1}\,e^{-k^{d}s}=\frac{1}{s\,d}(e^{-\lambda^{d}s})\,,
\ee  
up to irrelevant numerical factors.
In the above intergral we have introduced an IR cutoff $\lambda$, which has been chosen large enough that the UV approximation for $g_{reg}^{(d)}$ is still valid, i.e. $\lambda \gg l^{-1}$. Substituting this back into (\ref{spectral}), we find 
\be
d_{s}^{UV}= 2+2\lambda^{D}s\rightarrow 2\,, \qquad \text{as} \; s\rightarrow 0\,.
\label{spectraluv}
\ee 
This shows that the improved UV behaviour of the regularised nonlocal Green functions leads to dimensional reduction to $d_s=2$ for all spacetime dimensions considered. Furthermore, the linear behaviour in $s$ with large coefficient $\lambda^d$ is in accordance with the numerical evidence provided in Figure \ref{2}. It is important to note that (\ref{spectraluv}) is only valid for small values (with respect to the non-locality scale $l$) of the diffusion parameter $s\ll l$.

Finally, we should pay attention to the fact that the universal dimensional reduction to $d_s=2$, crucially relies on the regularisation of the Green functions. One might therefore ask what happens should one decide not to perform the regularisation? 
The first hint that something will go wrong with the computation of $d_s$, arises from the fact that $g^{(d)}(k^2)\rightarrow \text{constant}$ as $k^2\rightarrow 0$, so that the integral in (\ref{prob}) clearly diverges. To analyse the small scale behaviour of $d_s$ we must therefore introduce both an IR cutoff $\lambda$ and a UV cutoff $\Lambda$. Then 
\begin{align}
P(s)&\approx \lim_{\lambda\rightarrow 0,\,\Lambda\rightarrow\infty}\int_{\lambda}^{\Lambda}dk \,k^{d-1}e^{-c_{d}s}\nn
&= \lim_{\lambda\rightarrow 0,\,\Lambda\rightarrow\infty}e^{-c_{d}s}\frac{\Lambda^{d}-\lambda^{d}}{d},
\end{align}
where we used the fact that in the UV, $g^{(d)}$ goes like a dimensionful and dimension dependent constant, $c_{d}$, which therefore dominates the integral. Substituting this into (\ref{spectral}) and ignoring issues about the order of limits, we find that $d_{s}= 2c_{d} s$. This result is in accordance with the numerical simulations we have performed, and clearly leads to a spectral dimension that does not make much physical sense. Indeed, it does not asymptote the Hausdorff dimension $d$ for $s\gg l$, but actually diverges.

One could  have argued from the beginning that without regularisation the spectral dimension would be meaningless, and that it would possess a linear dependence in $s$ for all $s$. For instance, consider the spectral dimension in a spacetime with $d=2$. Then,
\begin{align} \label{4ds}
-2s\frac{\int _{0}^{\infty }dz\; g_{2}(z) e^{g_{2}(z) s}}{\int _{0}^{\infty }\text{dz}\; e^{g_{2}(z) s}}=4 \rho  s\,,
\end{align}
where $z=k^{2}$. To obtain \eqref{4ds} we split both the numerator and denominator by introducing an IR cutoff
$L$ big enough to approximate the integrand with its UV constant behaviour and use the fact that 
$\int _{0}^{\infty}\text{dz}\; e^{g_{2}(z) s}$ diverges.
The factor of 4 obtained in the final expression is specific to the minimal non-local d'Alembertian in $d=2$.     

\subsection{Numerical results} \label{numerical}

\begin{figure}[H]
\centering
\includegraphics[scale=0.3]{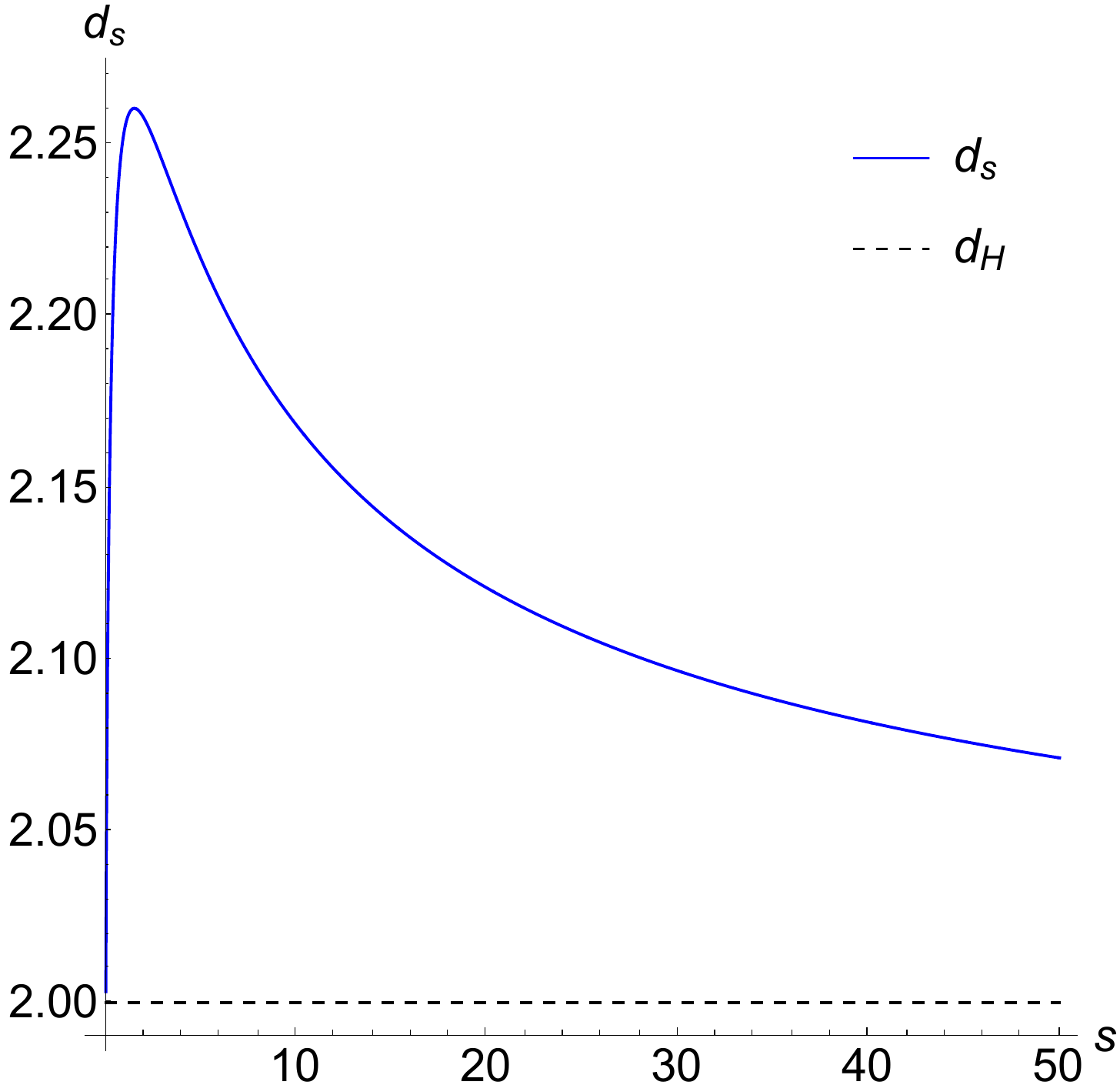}
\includegraphics[scale=0.3]{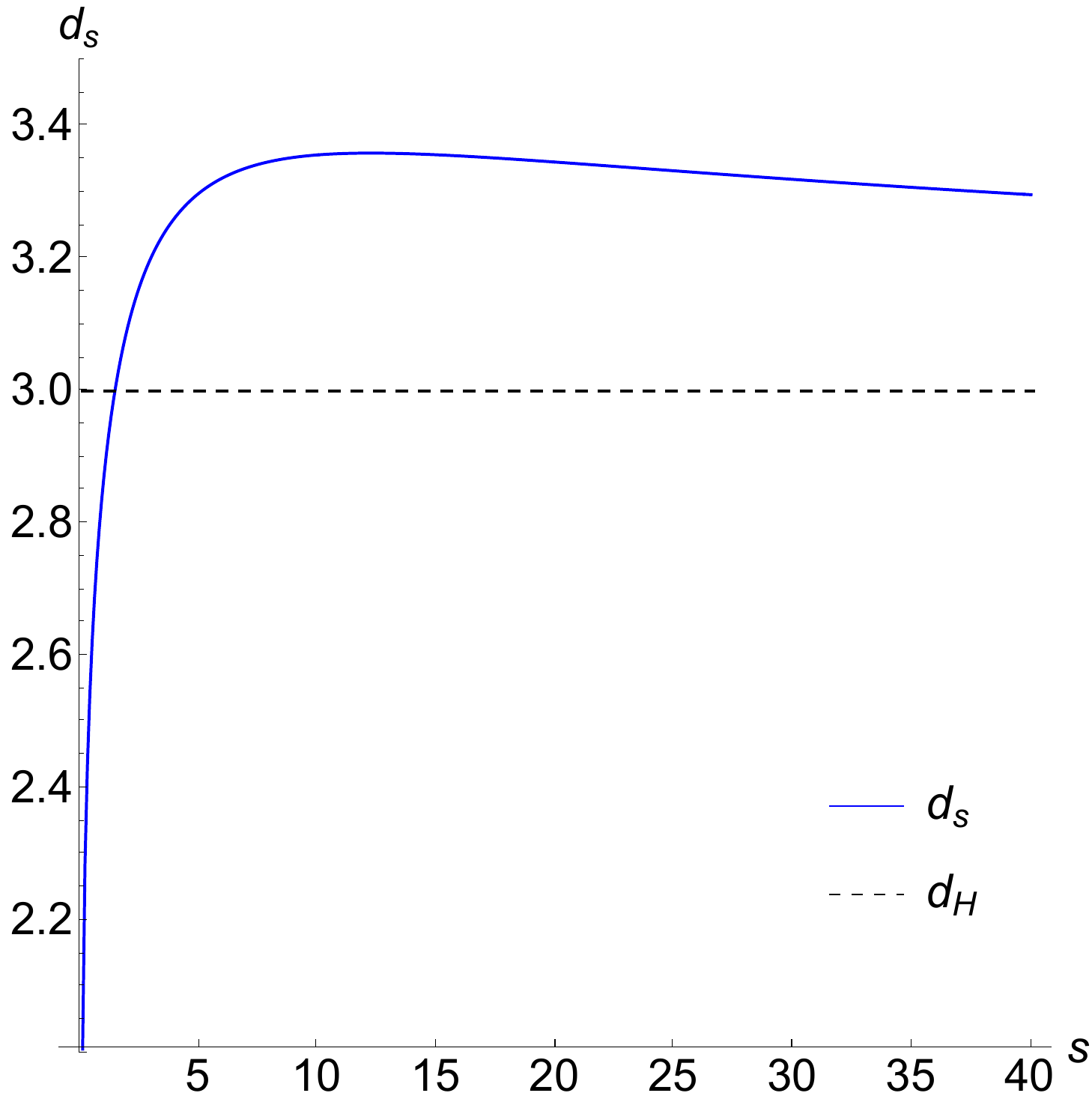}
\includegraphics[scale=0.3]{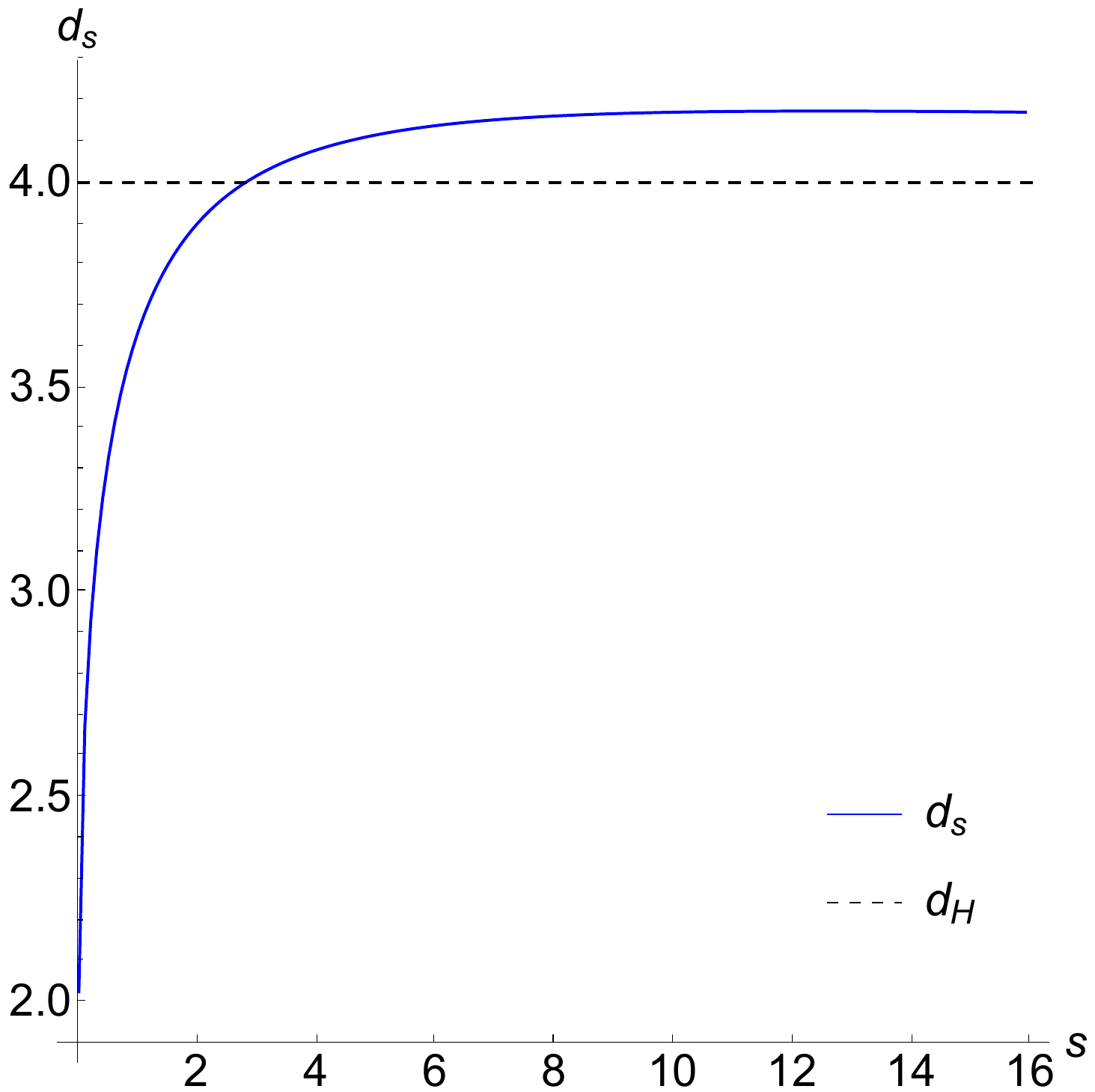}
\caption{From top to bottom we have the spectral dimension $d_s$ as a function of diffusion time $s$ for $\rho=1$ for $d=2,3$ and 4 respectively. The dashed line represents the value of the Hausdorff dimension. In the 2$d$ plot the spectral dimension can clearly be seen to interpolate between $d_s=2$ at short scales and $d_s=2$ at large scales, with a maximum occurring for $s\approx \rho^{1/2}$. The large scale asymptotic behaviour and the maximum are not as clear in the 3 and 4 dimensional cases due to poorer numerics in $d>2$. Nonetheless one can see the short scale limit to $d_s=2$, a maximum for $s\approx 10\rho^{1/d}$, and a large scale limit which is very slowly asymptoting the Hausdorff dimension. These numerical results corroborate the analytical analysis of Section \ref{ds} and provide evidence for the behaviour of $d_s$ interpolating between short and large scales}\label{2}
\end{figure}

We numerically calculate the spectral dimension in 2, 3 and 4 dimensions using the regularised minimal non-local d'Alembertians. These are given in terms of the unregularised d'Alembertians \cite{Aslanbeigi:2014tg}: 
\begin{widetext}
\begin{align}
g^{(2)}(k^{2})&=a^{(2)}\rho+2\rho\sum_{n=0}^{2}\frac{b_{n}^{(2)}}{n!}\left(\frac{\sqrt{\pi}}{4}\right)^{n}\int_{0}^{\infty}d\xi\;\xi^{2n+1}e^{-\frac{\sqrt{\pi}}{4}\xi^{2}}K_{0}\left(\sqrt{\frac{k^{2}}{\rho}}\;\xi\right),\\
g^{(3)}(k^{2})&=a^{(3)}\rho^{2/3}+2(2\pi)^{1/2}\rho^{5/6}(k^{2})^{-1/4}\sum_{n=0}^{2}\frac{b_{n}^{(3)}}{n!}\left(\frac{\pi}{12}\right)^{n}  \int_{0}^{\infty}d\xi\;\xi^{3n+3/2}e^{-\frac{\pi}{12}\xi^{3}}K_{1/2}\left(\sqrt{k^{2}}\rho^{-1/3}\xi\right),\\
g^{(4)}(k^{2})&=a^{(4)}\sqrt{\rho}+4\pi\left(\frac{k^{2}}{\sqrt{\rho}}\right)^{-1/2} \sum_{n=0}^{3}\frac{b_{n}^{(4)}}{n!}\left(\frac{\pi}{24}\right)^{n}\int_{0}^{\infty}d\xi\;\xi^{4n+2}e^{-\frac{\pi}{24}\xi^{4}}K_{1}\left(\sqrt{\frac{k^{2}}{\sqrt{\rho}}}\;\xi\right),
\end{align}
in $d=2,3$ and 4 respectively,
where $K_{n}$ are modified Bessel functions of second type and $$a^{(2)}=-2,\ b_{0}^{(2)}=4,\ b_{1}^{(2)}=-8,\ b_{2}^{(2)}=4,$$
$$a^{(3)}=-\frac{1}{\Gamma(5/3)}\left(\frac{\pi}{3\sqrt{2}}\right)^{2/3},\ b_{0}^{(3)}=\frac{1}{\Gamma(5/3)}\left(\frac{\pi}{3\sqrt{2}}\right)^{2/3},$$ $$b_{1}^{(3)}=-\frac{27}{16}\left(\frac{\pi}{2\sqrt{3}}\right)^{2/3}\frac{1}{\Gamma(2/3)},\ b_{2}^{(4)}=\frac{9}{8}\left(\frac{\pi}{\sqrt{6}}\right)^{2/3}\frac{1}{\Gamma(2/3)}.$$
$$a^{(4)}=\frac{-4}{\sqrt{6}},\ b_{0}^{(4)}=\frac{4}{\sqrt{6}},\ b_{1}^{(4)}=\frac{-36}{\sqrt{6}},\ b_{2}^{(4)}=\frac{64}{\sqrt{6}},\ b_{3}^{(4)}=\frac{-32}{\sqrt{6}},$$
\end{widetext}
as obtained via equation (\ref{greg}).
The results given in Figure \ref{2} show that in $d=2,3,4$ the spectral dimension approaches 2 in the limit $s\rightarrow0$,
increases to a maximum greater than the large scale Hausdorff dimension when $s\sim O(l)$, and then decays back to the
Hausdorff dimension as $s\rightarrow\infty$.
\footnote{Note that in $d=3,4$ the maximum occurs on scales $s\sim10l$ rather than $l$ and in the $s\rightarrow\infty$ 
the decay back to the Hausdorff dimension is not as marked as in 2$d$. We believe that these minor differences are due to
the higher degree of numerical approximation present in their analysis.}

\section{Discussion}\label{disc}

We have studied the spectral dimension arising from non-local d'Alembertians derived from causal-set theory. 
We have shown that after regularising these d'Alembertians by removing an unphysical
\footnote{From the point of view of the fundamental discrete theory, in which the coincidence limit is ill-defined.} 
divergence in the coincidence limit, dimensional reduction is present. In particular the spectral dimension goes to 2 
at short scales in every dimension. The small scale dimensional reduction can be seen to arise from the improved 
UV  behaviour of the propagator for such non-local theories. We have also provided numerical evidence for $d_s$ 
as a function of diffusion time $s$ for the minimal regularised d'Alembertians in $d=2,3$ and 4, confirming our analytical 
estimates in the limits $s\ll l$ and $s\gg l$. These simulations show that in all three cases considered the spectral dimension possesses a maximum on scales of order $l$, and asymptotes the Hausdorff dimension from above. This is an indication of super-diffusive behaviour on intermediate scales --- where effects due to the non-locality scale start to become relevant --- a feature similar to that found in \cite{Sotiriou:2011mu} in the context of causal dynamical triangulations. 

\section{Conclusions}

To conclude with we would like to speculate on the significance and validity of these results. Recall first that the UV behaviour  $g^{(d)}_{reg}\rightarrow (k^2)^{d/2}$ was key in ensuring that $d_s(s)\rightarrow 2$ as $s\rightarrow0$ in all dimensions. Now, it was shown in \cite{Aslanbeigi:2014tg} and \cite{Belenchia:2014fda}, where the quantum theory of free scalar fields with nonlocal dynamics was studied, that these operators lead to Wightman functions of the form
\be
 W(x,y)  = \!\! \int_0^\infty \!\!\!\! d\mu^2 \rho(\mu^2) \!\! \int \!  \fr{d^dk 2 \pi }{(2\pi)^{d}} \theta(k^0)\delta(k^2+\mu^2)e^{ik\cdot (x-y)},
\nonumber 
\ee
where $\rho(-k^2)= \text{Im}(g^{(d)}(k^2))/|g^{(d)}(k^2)|^2$, a form reminiscent of the K\"allen-Lehman representation of the Wightman function in interacting QFTs.\footnote{We have ignored contributions coming from complex mass poles in the propagator since they are not relevant for the current discussion and their inclusion does not change the conclusion of this argument.} Unlike the  K\"allen-Lehman representation of local interacting QFTs, however, the spectral function $\rho$ arising from the causet derived nonlocal d'Alembertians defined in \cite{Aslanbeigi:2014tg} is not a positive function in general. In fact, evidence suggests that for $d>2$ all operators of this kind will lead to a non-positive spectral functions $\rho$. This implies the existence of negative norm states in the quantum description of these theories.

As was explicitly stated by Weinberg (see Ref.~\cite{Weinberg:1996kr} Section 10.7, p.460), a consequence of the  K\"allen-Lehman spectral representation, and the positivity of the spectral function $\rho(\mu^2)$ (which in standard local QFT is given by the sum of squares of matrix elements of the field-observable), is that the propagator cannot vanish for $|k^2|\rightarrow\infty$ faster than the bare propagator $1/k^2$. Since the Wightman function of nonlocal theories of the kind studied in this paper can be written {\it \`a la} K\"allen-Lehman, a natural question is whether the improved UV behaviour of the propagator, and therefore the small scale dimensional reduction in the spectral dimension, are inextricably linked to negative norm states in the quantum theory. If this were the case, it would raise doubts as to the physical relevance of our result in the regime $s\ll l$.

Knowing about the issue of non-positive spectral functions associated to causet derived nonlocal d'Alembertians, Aslanbeigi and Saravani studied QFTs of generalised versions of the operators in \cite{Aslanbeigi:2014tg}, where they imposed the positivity of $\rho(\mu^2)$ from the beginning \cite{Saravani:2015aa}. Hence, from Weinberg's argument one would expect  that their newly defined $\tilde{g}^{(4)}_{reg}(k^2)$ goes like $k^2$ as $k^2\rightarrow\infty$, which is indeed the case.
Because of this the spectral dimension does not reduce to 2 at small scales but rather starts off at 4, increases, and then asymptotes back to 4 from above at large scales (much like the 2$d$ case considered in Section \ref{numerical}). Although interesting in its own right, the specific theory they consider does not fall within the family of d'Alembertians used in this paper, and as such cannot be (at least trivially) seen as continuum approximations to the more fundamental operators living on an underlying causal set, in the way Sorkin had originally constructed them.  

A possible way out of these issues arising in the quantum theory is to take the non-locality scale $l$ to correspond to the causet fundamental discreteness scale. In this case one would expect that on scales $s\sim l$ the diffusing particle would cease to see a continuum spacetime, and would start feeling the underlying discreteness of the causal set. Therefore, at these scales the analysis performed in this paper would cease to be applicable. In order  to study the behaviour of the spectral dimension on scales $s< l$, one would instead  have to resort to methods to be deployed directly to the causal set itself, e.g. Eichhorn {\it et al.}'s method \cite{Eichhorn:2013ova}. We conclude by noting that our analysis shows an increasing $d_s(s)$ as $s\rightarrow l^+$, which then only starts to decrease after it has gone through $s\sim O(l)$. This might be a hint of the existence of a universal description of the spectral dimension which interpolates between our results and the EM spectral dimension. 

\mbox{}

\section*{Note added}

After this work was completed the article~\cite{Carlip:2015mra}~by S.~Carlip  appeared in the electronic archives with similar results.  

\acknowledgments 
AB and DMTB  would like to thank Stefano Liberati and Marco Letizia for useful discussions during the development of this project and acknowledge financial support from the John Templeton Foundation (JTF), grant  No. 51876. AM wish to acknowledge support by the Shanghai Municipality, through the grant No. KBH1512299, and by Fudan University, through the grant No. JJH1512105.

\bibliographystyle{apsrev}
\bibliography{refs}

\begin{thebibliography}{26}
\expandafter\ifx\csname natexlab\endcsname\relax\def\natexlab#1{#1}\fi
\expandafter\ifx\csname bibnamefont\endcsname\relax
  \def\bibnamefont#1{#1}\fi
\expandafter\ifx\csname bibfnamefont\endcsname\relax
  \def\bibfnamefont#1{#1}\fi
\expandafter\ifx\csname citenamefont\endcsname\relax
  \def\citenamefont#1{#1}\fi
\expandafter\ifx\csname url\endcsname\relax
  \def\url#1{\texttt{#1}}\fi
\expandafter\ifx\csname urlprefix\endcsname\relax\def\urlprefix{URL }\fi
\providecommand{\bibinfo}[2]{#2}
\providecommand{\eprint}[2][]{\url{#2}}

\bibitem[{\citenamefont{Ambjorn et~al.}(2005)\citenamefont{Ambjorn, Jurkiewicz,
  and Loll}}]{Ambjorn:2005db}
\bibinfo{author}{\bibfnamefont{J.}~\bibnamefont{Ambjorn}},
  \bibinfo{author}{\bibfnamefont{J.}~\bibnamefont{Jurkiewicz}},
  \bibnamefont{and} \bibinfo{author}{\bibfnamefont{R.}~\bibnamefont{Loll}},
  \bibinfo{journal}{Phys. Rev. Lett.} \textbf{\bibinfo{volume}{95}},
  \bibinfo{pages}{171301} (\bibinfo{year}{2005}), \eprint{hep-th/0505113}.

\bibitem[{\citenamefont{Lauscher and Reuter}(2005)}]{Lauscher:2005qz}
\bibinfo{author}{\bibfnamefont{O.}~\bibnamefont{Lauscher}} \bibnamefont{and}
  \bibinfo{author}{\bibfnamefont{M.}~\bibnamefont{Reuter}},
  \bibinfo{journal}{JHEP} \textbf{\bibinfo{volume}{0510}}, \bibinfo{pages}{050}
  (\bibinfo{year}{2005}), \eprint{hep-th/0508202}.

\bibitem[{\citenamefont{Modesto}(2009)}]{Modesto:2008jz}
\bibinfo{author}{\bibfnamefont{L.}~\bibnamefont{Modesto}},
  \bibinfo{journal}{Class.Quant.Grav.} \textbf{\bibinfo{volume}{26}},
  \bibinfo{pages}{242002} (\bibinfo{year}{2009}), \eprint{0812.2214}.

\bibitem[{\citenamefont{Modesto}(2012{\natexlab{a}})}]{Modesto:2011kw}
\bibinfo{author}{\bibfnamefont{L.}~\bibnamefont{Modesto}},
  \bibinfo{journal}{Phys. Rev.} \textbf{\bibinfo{volume}{D86}},
  \bibinfo{pages}{044005} (\bibinfo{year}{2012}{\natexlab{a}}),
  \eprint{1107.2403}.

\bibitem[{\citenamefont{Modesto}(2012{\natexlab{b}})}]{Modesto:2012ys}
\bibinfo{author}{\bibfnamefont{L.}~\bibnamefont{Modesto}}
  (\bibinfo{year}{2012}{\natexlab{b}}), \eprint{1202.3151}.

\bibitem[{\citenamefont{Modesto and Rachwal}(2014)}]{Modesto:2014lga}
\bibinfo{author}{\bibfnamefont{L.}~\bibnamefont{Modesto}} \bibnamefont{and}
  \bibinfo{author}{\bibfnamefont{L.}~\bibnamefont{Rachwal}},
  \bibinfo{journal}{Nucl. Phys.} \textbf{\bibinfo{volume}{B889}},
  \bibinfo{pages}{228} (\bibinfo{year}{2014}), \eprint{1407.8036}.

\bibitem[{\citenamefont{Benedetti}(2009)}]{Benedetti:2008gu}
\bibinfo{author}{\bibfnamefont{D.}~\bibnamefont{Benedetti}},
  \bibinfo{journal}{Phys.Rev.Lett.} \textbf{\bibinfo{volume}{102}},
  \bibinfo{pages}{111303} (\bibinfo{year}{2009}), \eprint{0811.1396}.

\bibitem[{\citenamefont{Arzano and Trzesniewski}(2014)}]{Arzano:2014jfa}
\bibinfo{author}{\bibfnamefont{M.}~\bibnamefont{Arzano}} \bibnamefont{and}
  \bibinfo{author}{\bibfnamefont{T.}~\bibnamefont{Trzesniewski}},
  \bibinfo{journal}{Phys.Rev.} \textbf{\bibinfo{volume}{D89}},
  \bibinfo{pages}{124024} (\bibinfo{year}{2014}), \eprint{1404.4762}.

\bibitem[{\citenamefont{Calcagni}(2009)}]{Calcagni:2009ar}
\bibinfo{author}{\bibfnamefont{G.}~\bibnamefont{Calcagni}},
  \bibinfo{journal}{JHEP} \textbf{\bibinfo{volume}{0909}}, \bibinfo{pages}{112}
  (\bibinfo{year}{2009}), \eprint{0904.0829}.

\bibitem[{\citenamefont{Calcagni et~al.}(2014)\citenamefont{Calcagni, Modesto,
  and Nardelli}}]{Calcagni:2014wba}
\bibinfo{author}{\bibfnamefont{G.}~\bibnamefont{Calcagni}},
  \bibinfo{author}{\bibfnamefont{L.}~\bibnamefont{Modesto}}, \bibnamefont{and}
  \bibinfo{author}{\bibfnamefont{G.}~\bibnamefont{Nardelli}}
  (\bibinfo{year}{2014}), \eprint{1408.0199}.

\bibitem[{\citenamefont{Carlip}(2009)}]{Carlip:2009km}
\bibinfo{author}{\bibfnamefont{S.}~\bibnamefont{Carlip}}, pp.
  \bibinfo{pages}{69--84} (\bibinfo{year}{2009}), \eprint{1009.1136}.

\bibitem[{\citenamefont{Eichhorn and Mizera}(2014)}]{Eichhorn:2013ova}
\bibinfo{author}{\bibfnamefont{A.}~\bibnamefont{Eichhorn}} \bibnamefont{and}
  \bibinfo{author}{\bibfnamefont{S.}~\bibnamefont{Mizera}},
  \bibinfo{journal}{Class.Quant.Grav.} \textbf{\bibinfo{volume}{31}},
  \bibinfo{pages}{125007} (\bibinfo{year}{2014}), \eprint{1311.2530}.

\bibitem[{\citenamefont{Benincasa and Dowker}(2010)}]{Benincasa:2010ac}
\bibinfo{author}{\bibfnamefont{D.~M.~T.} \bibnamefont{Benincasa}}
  \bibnamefont{and} \bibinfo{author}{\bibfnamefont{F.}~\bibnamefont{Dowker}},
  \bibinfo{journal}{Phys. Rev. Lett.} \textbf{\bibinfo{volume}{104}},
  \bibinfo{pages}{181301} (\bibinfo{year}{2010}),
  \eprint{http://arxiv.org/abs/1001.2725}.

\bibitem[{\citenamefont{Sorkin}(2006)}]{Sorkin:2007qi}
\bibinfo{author}{\bibfnamefont{R.~D.} \bibnamefont{Sorkin}}, in
  \emph{\bibinfo{booktitle}{{Approaches to Quantum Gravity: Towards a New
  Understanding of Space and Time}}}, edited by
  \bibinfo{editor}{\bibfnamefont{D.}~\bibnamefont{Oriti}}
  (\bibinfo{publisher}{Cambridge University Press}, \bibinfo{year}{2006}),
  \eprint{gr-qc/0703099}.

\bibitem[{\citenamefont{Aslanbeigi et~al.}(2014)\citenamefont{Aslanbeigi,
  Saravani, and Sorkin}}]{Aslanbeigi:2014tg}
\bibinfo{author}{\bibfnamefont{S.}~\bibnamefont{Aslanbeigi}},
  \bibinfo{author}{\bibfnamefont{M.}~\bibnamefont{Saravani}}, \bibnamefont{and}
  \bibinfo{author}{\bibfnamefont{R.~D.} \bibnamefont{Sorkin}},
  \bibinfo{journal}{JHEP} \textbf{\bibinfo{volume}{1406}}, \bibinfo{pages}{024}
  (\bibinfo{year}{2014}), \eprint{1403.1622},
  \urlprefix\url{http://arxiv.org/abs/1403.1622}.

\bibitem[{\citenamefont{Bombelli et~al.}(1987)\citenamefont{Bombelli, Lee,
  Meyer, and Sorkin}}]{Bombelli:1987aa}
\bibinfo{author}{\bibfnamefont{L.}~\bibnamefont{Bombelli}},
  \bibinfo{author}{\bibfnamefont{J.-H.} \bibnamefont{Lee}},
  \bibinfo{author}{\bibfnamefont{D.}~\bibnamefont{Meyer}}, \bibnamefont{and}
  \bibinfo{author}{\bibfnamefont{R.}~\bibnamefont{Sorkin}},
  \bibinfo{journal}{Phys. Rev. Lett} \textbf{\bibinfo{volume}{59}},
  \bibinfo{pages}{521} (\bibinfo{year}{1987}).

\bibitem[{\citenamefont{Bombelli et~al.}(2009)\citenamefont{Bombelli, Henson,
  and Sorkin}}]{Bombelli:2006nm}
\bibinfo{author}{\bibfnamefont{L.}~\bibnamefont{Bombelli}},
  \bibinfo{author}{\bibfnamefont{J.}~\bibnamefont{Henson}}, \bibnamefont{and}
  \bibinfo{author}{\bibfnamefont{R.~D.} \bibnamefont{Sorkin}},
  \bibinfo{journal}{Mod. Phys. Lett.} \textbf{\bibinfo{volume}{A24}},
  \bibinfo{pages}{2579} (\bibinfo{year}{2009}), \eprint{gr-qc/0605006}.

\bibitem[{\citenamefont{Dowker and Glaser}(2013)}]{Dowker:2013vl}
\bibinfo{author}{\bibfnamefont{F.}~\bibnamefont{Dowker}} \bibnamefont{and}
  \bibinfo{author}{\bibfnamefont{L.}~\bibnamefont{Glaser}},
  \bibinfo{journal}{2013,Class. Quantum Grav. 30 195016}
  (\bibinfo{year}{2013}), \eprint{1305.2588},
  \urlprefix\url{http://arxiv.org/abs/1305.2588}.

\bibitem[{\citenamefont{Glaser}(2013)}]{Glaser:2013sf}
\bibinfo{author}{\bibfnamefont{L.}~\bibnamefont{Glaser}}
  (\bibinfo{year}{2013}), \eprint{1311.1701},
  \urlprefix\url{http://arxiv.org/abs/1311.1701}.

\bibitem[{\citenamefont{Dom{\'\i}nguez and
  Trione}(1979)}]{dominguez1979laplace}
\bibinfo{author}{\bibfnamefont{A.~G.} \bibnamefont{Dom{\'\i}nguez}}
  \bibnamefont{and} \bibinfo{author}{\bibfnamefont{S.~E.}
  \bibnamefont{Trione}}, \bibinfo{journal}{Advances in Mathematics}
  \textbf{\bibinfo{volume}{31}}, \bibinfo{pages}{51} (\bibinfo{year}{1979}).

\bibitem[{\citenamefont{Saravani and Aslanbeigi}(2015)}]{Saravani:2015aa}
\bibinfo{author}{\bibfnamefont{M.}~\bibnamefont{Saravani}} \bibnamefont{and}
  \bibinfo{author}{\bibfnamefont{S.}~\bibnamefont{Aslanbeigi}}
  (\bibinfo{year}{2015}), \eprint{1502.01655},
  \urlprefix\url{http://arxiv.org/abs/1502.01655}.

\bibitem[{\citenamefont{Belenchia et~al.}(2015)\citenamefont{Belenchia,
  Benincasa, and Liberati}}]{Belenchia:2014fda}
\bibinfo{author}{\bibfnamefont{A.}~\bibnamefont{Belenchia}},
  \bibinfo{author}{\bibfnamefont{D.~M.~T.} \bibnamefont{Benincasa}},
  \bibnamefont{and} \bibinfo{author}{\bibfnamefont{S.}~\bibnamefont{Liberati}},
  \bibinfo{journal}{JHEP} \textbf{\bibinfo{volume}{1503}}, \bibinfo{pages}{036}
  (\bibinfo{year}{2015}), \eprint{1411.6513}.

\bibitem[{\citenamefont{Benedetti and Henson}(2009)}]{Benedetti:2009ge}
\bibinfo{author}{\bibfnamefont{D.}~\bibnamefont{Benedetti}} \bibnamefont{and}
  \bibinfo{author}{\bibfnamefont{J.}~\bibnamefont{Henson}},
  \bibinfo{journal}{Phys. Rev.} \textbf{\bibinfo{volume}{D80}},
  \bibinfo{pages}{124036} (\bibinfo{year}{2009}), \eprint{0911.0401}.

\bibitem[{\citenamefont{Sotiriou et~al.}(2011)\citenamefont{Sotiriou, Visser,
  and Weinfurtner}}]{Sotiriou:2011mu}
\bibinfo{author}{\bibfnamefont{T.~P.} \bibnamefont{Sotiriou}},
  \bibinfo{author}{\bibfnamefont{M.}~\bibnamefont{Visser}}, \bibnamefont{and}
  \bibinfo{author}{\bibfnamefont{S.}~\bibnamefont{Weinfurtner}},
  \bibinfo{journal}{Phys.Rev.Lett.} \textbf{\bibinfo{volume}{107}},
  \bibinfo{pages}{131303} (\bibinfo{year}{2011}), \eprint{1105.5646}.

\bibitem[{\citenamefont{Weinberg}(1996)}]{Weinberg:1996kr}
\bibinfo{author}{\bibfnamefont{S.}~\bibnamefont{Weinberg}}
  (\bibinfo{year}{1996}).

\bibitem[{\citenamefont{Carlip}(2015)}]{Carlip:2015mra}
\bibinfo{author}{\bibfnamefont{S.}~\bibnamefont{Carlip}}
  (\bibinfo{year}{2015}), \eprint{1506.08775}.

\end{thebibliography}

\end{document}